\documentclass[aps,pre,groupedaddress,reprint]{revtex4-1}

\usepackage{graphicx}

\begin{document}


\title{Cities as nuclei of sustainability?}


\author{Diego Rybski}
\email[]{ca-dr@rybski.de}
\affiliation{Potsdam Institute for Climate Impact Research, 
14412 Potsdam, Germany, EU}

\author{Dominik E. Reusser}
\affiliation{Potsdam Institute for Climate Impact Research, 
14412 Potsdam, Germany, EU}

\author{Anna-Lena Winz}
\affiliation{Potsdam Institute for Climate Impact Research, 
14412 Potsdam, Germany, EU}

\author{Christina Fichtner}
\affiliation{Potsdam Institute for Climate Impact Research, 
14412 Potsdam, Germany, EU}

\author{Till Sterzel}
\affiliation{Potsdam Institute for Climate Impact Research, 
14412 Potsdam, Germany, EU}

\author{J\"urgen P. Kropp}
\affiliation{Potsdam Institute for Climate Impact Research, 
14412 Potsdam, Germany, EU}
\affiliation{University of Potsdam, Dept. of Geo- \& Environmental Sciences, 
14476 Potsdam, Germany, EU}


\date{\today}

\begin{abstract}
We have assembled CO$_2$ emission figures from collections of urban GHG emission estimates published in peer reviewed journals or reports from research institutes and non-governmental organizations.
Analyzing the scaling with population size we find that the exponent is development dependent with a transition from super- to sub-linear scaling.
From the climate change mitigation point of view, the results suggest that urbanization is desirable in developed countries and should be avoided in developing ones. 
Further, we compare this analysis with a second scaling relation, namely the fundamental allometry between city population and area, and propose that density might be the decisive quantity.
Last, we derive the theoretical country-wide urban emissions by integration and obtain a dependence on the size of the largest city.
\end{abstract}


\maketitle

\section{Introduction}
\label{sec:intro}

Already in the first half of the 20th century G.\,K.~Zipf thought about how urban indicators scale with the city size and finds e.g.\ proportionality between the number of personal \& business service establishments and the population \cite{ZipfGK1949}. 
More than half a century later, inspired by the analogy of the city as a metabolism, \citeauthor{BettencourtLHKW2007}~\citeyear{BettencourtLHKW2007}~\cite{BettencourtLHKW2007} analyze various indicators and measured super-linear scaling for quantities related to innovation or wealth, linear scaling for quantities of individual human needs, and sub-linear scaling for material quantities and infrastructure. 
Exploring these results, the authors present a city model based on the concept of carrying capacity that leads to alternating phases of critical growth and collapse \cite{BettencourtLHKW2007}.

Meanwhile, urban scaling has been investigated for several other city properties, including urban green space \cite{FullerG2009}; prosocial behavior \cite{ArbesmanC2011}; homicides \cite{GomezLievanoYB2012}; child labor, elderly population, literacy, unemployment \citep{AlvesRLM2013}; bank card transactions \cite{SobolevskySGCHAR2014}; and agriculture \cite{DAutiliaDA2015}.
\citeauthor{AlvesMLR2015}~\citeyear{AlvesMLR2015}~\cite{AlvesMLR2015} employ a scale-adjusted metric that takes into account allometry.
The super-linear scaling, particularly of economic quantities, however, remains a mostly unsolved riddle and seems to have the character of a phenomenological law.

Different models based on human interactions and social networks have been proposed \cite{ArbesmanKS2009,SchlapferBGRCSWR2014,YakuboSK2014}, exploring a surplus stemming from the social network, i.e.\ more intense or frequent social interactions in large cities.
Despite a growing body of literature addressing the origins of non-linear urban scaling (e.g.\ \cite{BettencourtW2010,BettencourtLMA2013}), little attention has been given to the implications of accelerated socio-economic activity in cities.
In a more general sense, \citeauthor{BettencourtW2010}~\citeyear{BettencourtW2010}~\cite{BettencourtW2010} stated: \emph{"The many problems associated with urban growth and global sustainability, however, are typically treated as independent issues"}.

Especially in the global context, the question of carbon dioxide (CO$_2$) emissions from cities is of interest \cite{SatterthwaiteD2008,DodmanD2009,KennedySGHHHPPRM2009,GlaeserK2010,SovacoolB2010,MakidoDY2012,TollefsonJ2012,BereitschaftD2013,MarcotullioSASG2013,MinxBWBCFFPWH2013,OuLLC2013,JonesK2014,BaurLFK2015,CreutzigBBPS2015,LankaoNT2008} -- 
to what extent is urbanization driving or mitigating climate change? 

The question of whether small or large cities are more efficient in terms of per capita CO$_2$ emissions has been addressed by a few publications.
\citeauthor{FragkiasLSS2013}~\citeyear{FragkiasLSS2013}~\cite{FragkiasLSS2013} find that CO$_2$ emissions of U.S.\ metropolitan areas scale proportionally with urban population size.
In contrast, \citeauthor{OliveiraAM2014}~\citeyear{OliveiraAM2014}~\cite{OliveiraAM2014} report strong super-linear scaling for U.S.\ cities.
The authors attribute this discrepancy to the differing underlying spatial units, i.e.\ linearity is recovered for Metropolitan Statistical Areas (MSA) and super-linear scaling is obtained for cities defined as connected urban space \cite{OliveiraAM2014}.

\citeauthor{ArcauteHFYJB2014}~\citeyear{ArcauteHFYJB2014}~\cite{ArcauteHFYJB2014} define cities in England and Wales using commuting to work and population density thresholds in order to study a large set of urban indicators. 
The authors report that most urban indicators scale linearly with city size and conclude that population size alone does not provide sufficient information (see also \cite{CottineauHAB2015}).
Another perspective on various urban indicators is given by \citeauthor{LoufB2014SciRep}~\citeyear{LoufB2014SciRep}~\cite{LoufB2014SciRep},  who propose a model of urban growth and investigate the role of congestion \cite{LoufB2014SciRep,LoufB2014EPB}. 
Theoretically and empirically, the authors obtain super-linear scaling for transport-related CO$_2$ emissions.
Based on street network analysis, \citeauthor{MohajeriGF2015}~\citeyear{MohajeriGF2015}~\cite{MohajeriGF2015} also report super-linear scaling of transport CO$_2$ emissions with city population in Great Britain.

The work in hand follows a complementary approach. 
As a basic meta-study, compiled from various literature sources, it has the advantage of exhibiting better global coverage than previous studies. 
Grouping the cities according to the Gross Domestic Product (GDP) per capita of the corresponding countries, the results indicate non-universality since the scaling of CO$_2$ emissions versus population size depends on the economic development. 
While cities in economically less developed countries seem to exhibit super-linear scaling, cities in developed countries exhibit linearity or sub-linearity in respect to CO$_2$ emissions. 
Interestingly, another scaling, namely between population and area, resembles a similar picture when plotted as a function of time. 
Combining both scaling relations suggests that city density might play a decisive role.
Last we discuss theoretically how the total urban emissions of a country relate to the scaling of CO$_2$ emissions versus population size, and show that the population size of the largest city non-trivially influences the country emissions.
Empirically, we find that the population size of the largest city also follows a scaling relation with the population of the corresponding countries.

\section{Data}
\label{sec:data}

CO$_2$ emission figures were retrieved from collections of urban
GHG emission estimates published in peer reviewed journals or reports
from research institutes and non-governmental organizations.
We have converted all figures to CO$_2$ \emph{equivalent} and use the notation 
CO$_2$ for simplicity. 
Moreover, all values refer to \emph{annual} figures, again we omit "annual" for better readability.
The resulting set includes information on per capita urban GHG emission for 256 cities.
Published figures were largely drawn from primary sources such as city
assessments covering emissions in their territories. 
Figures were traced back to their original publication source for 
verification when necessary. 
The publications 
\cite{BrownSS2008},
\cite{BrownSS2009},
\cite{HoornwegSG2011}, 
\cite{KennedyRCD2009}, 
\cite{SovacoolB2010}, 
\cite{CarbonTrust2007},
\cite{Dore2006}, and 
\cite{ICLEI2009}
account for 246 of the figures (96\%).

The considered per capita emissions range from $0.1$\,t/cap for Rajshahi, Bangladesh \cite{ICLEI2009} to $43.7$\,t/cap for Aberystwyth, UK (derived from \cite{Dore2006} and \cite{CarbonTrust2007}). 
For 23 cities, both city and metropolitan region figures were
considered, as in general they are not identical \cite{HoornwegSG2011}.
Since this is a basic meta-study, that relies on previously published work, we cannot exclude that incoherent GHG methodologies, sources, sectors, accounting, and spatial units have been used.
By pooling the data, differences should average out, assuming the methods are evenly distributed among the pooled numbers.
When total emissions were not available, per cap emissions were multiplied by population figures from various sources such as United Nations World Urbanization Prospects, EUROSTAT, citypopulation.de, etc.

\section{City emissions}
\label{sec:cityemissions}

\begin{figure*}
\begin{centering}
\includegraphics[width=0.75\textwidth]{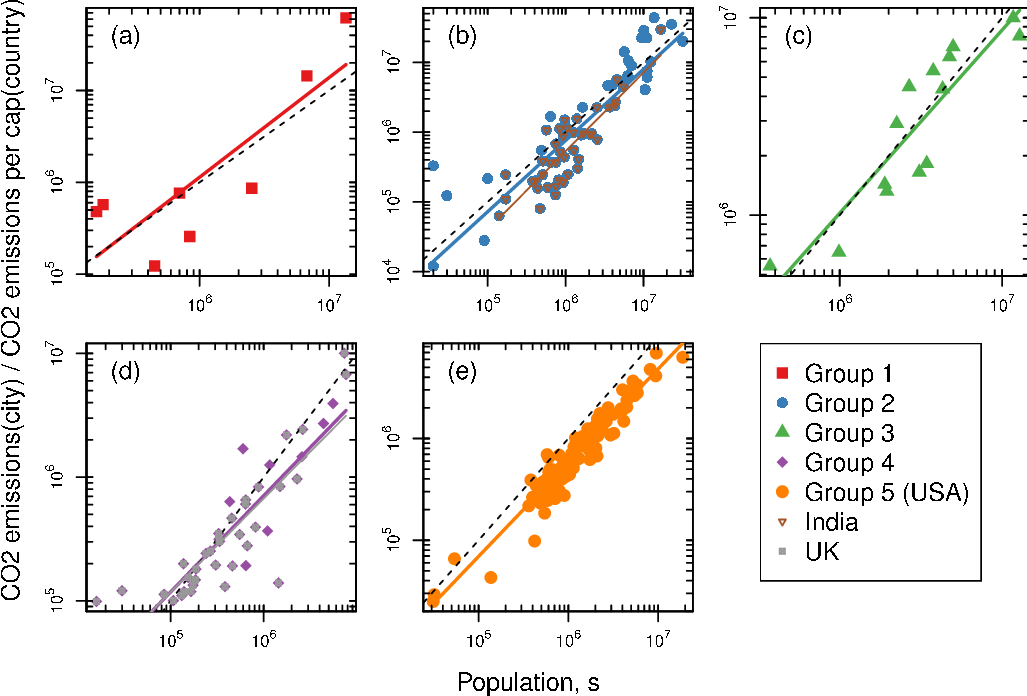}
\caption{
City efficiency in terms of CO$_2$ emissions.
Per capita (annual) CO$_2$ emissions divided by corresponding country emissions versus population and regressions according to Eq.~(\ref{eq:easg}).
The emission values have been normalized by dividing by the country 
emissions per capita.
City emission figures have been sorted and separated into groups 
according to the GDP/cap of the countries.
(a) Group~1 ($R^2=0.65$): Philippines (1), Bangladesh (4), Nepal (3).
(b) Group~2 ($R^2=0.78$): China (5), South Africa (1), Thailand (2), Greece (1), Sri Lanka (4), Slovenia (1), Czech Republic (1), Portugal (1), Spain (2), 
Mexico (1), South Korea (2), Indonesia (1), Brazil (3), India (42), 
Bhutan (2) and India (42, brown open triangles).
(c) Group~3 ($R^2=0.79$): Germany (4), Italy (4), Singapore (1), Belgium (1), Finland (1), 
Japan (1), France (1), Sweden (1).
(d) Group~4 ($R^2=0.75$): UK (35), Netherlands (1), Canada (4), Australia (1), 
Switzerland (1), Norway (1) and UK (35, grey open squares).
(e) Group~5 ($R^2=0.92$): USA (122). 
UK and India appear twice, as part of a group and as individual country.
Source of GDP data: \cite{Worldbank2013c}.
}
\label{fig:emissions}
\end{centering}
\end{figure*}

\begin{figure}
\begin{centering}
\includegraphics[width=\columnwidth]{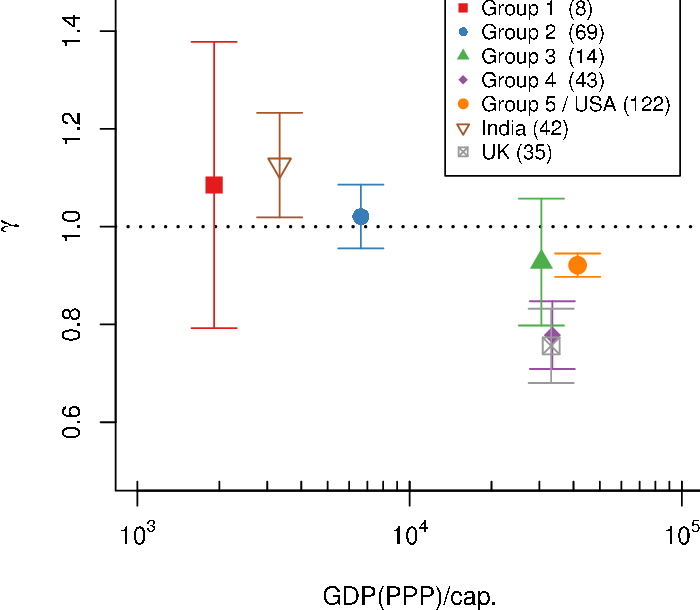}
\caption{
Economic development and city efficiency in terms of CO$_2$ emissions.
Determined exponents $\gamma$ from Eq.~(\ref{eq:easg}) 
versus the corresponding GDP/cap. 
The obtained power-law exponents $\gamma$ are plotted against the weighted 
average (according to the number of cities, see text) of the country's GDP/cap 
for the same groups and countries as in Fig.~\ref{fig:emissions}. 
The error-bars indicate the standard errors from fitting.
}
\label{fig:slopes}
\end{centering}
\end{figure}

We begin by grouping our database of 256 urban emission figures according to the country's GDP/cap, i.e.\ a balanced number of cases in each of five groups 
(all considered cities of a country are in the same group).
Based on these groups, we analyze the values by plotting the amount of CO$_2$ emissions $e$ against the corresponding city population $s$. 
Dividing by the per capita emissions of the corresponding 
countries we make the city emissions of different countries 
comparable. 
This step is only done for normalization reasons and would not be necessary if we had sufficient values for the various countries.

In Fig.~\ref{fig:emissions} we see correlations that approximately 
follow power-laws according to
\begin{equation}
e = d\,s^\gamma
\, ,
\label{eq:easg}
\end{equation}
where $d$ is a proportionality constant. 
For the linear case ($\gamma=1$) it is identical to the per capita emissions. 
The exponent $\gamma$ determines how strongly the emissions increase with city
size:
\begin{itemize}
\item $\gamma=1$ (linear) the larger the population, the higher the emissions; 
\item $\gamma<1$ (sub-linear) large cities emit less CO$_2$ per capita compared to small cities; and 
\item $\gamma>1$ (super-linear) large cities emit more CO$_2$ per capita
compared to small ones.
\end{itemize}
In addition to the evaluation by the five country groups, we
also estimated $\gamma$ for the three countries with more than 10 cities
in our database: India (part of group~2), UK
(part of group~4) and USA (constituting all of
group~5). 
Interestingly, our findings for the USA are consistent with the results reported in \cite{FragkiasLSS2013}, where a $\gamma$ slightly below $1$ is found.

The obtained exponents $\gamma$ for each group or country 
are plotted against the
corresponding GDP/cap values in Fig.~\ref{fig:slopes}. 
For groups with more than one country
a weighted country GDP was calculated based
on the number of cities in each country in
the group.
The Figure suggests that countries with 
lower GDP/cap values tend to exhibit $\gamma>1$ and 
countries with higher GDP/cap values tend to exhibit 
smaller $\gamma$, below or close to $1$.
Notwithstanding the large error-bars, the results indicate that there is a 
difference in scaling between developing and developed countries, 
i.e.\ cities in (economically) developing countries 
tend to a super-linear relation between population and
CO$_2$ emissions ($\gamma>1$), 
while cities of (economically) developed countries 
tend to a sub-linear relation ($\gamma<1$).
This suggests a transition at $\gamma\simeq 1$. 
In other words, large cities seem to be more \emph{CO$_2$ efficient} 
in economically stronger countries and 
more \emph{inefficient} in economically weaker countries.

\begin{figure}
\begin{centering}
\includegraphics[width=\columnwidth]{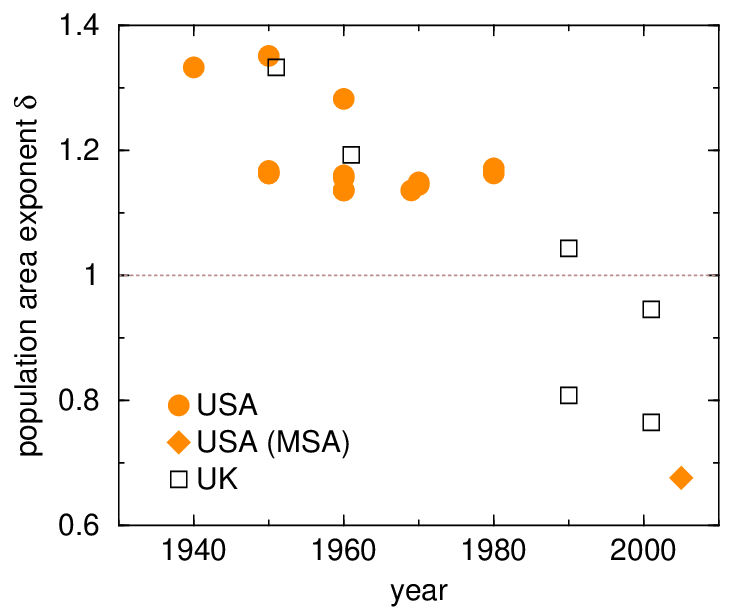}
\caption{
Exponents of the power-law relating population and area, Eq.~(\ref{eq:sadelta}), as obtained from the meta-study \cite{BattyF2011}. 
The 2005 value of the US is based on MSA.
The exponent~$\delta$ is plotted as a function of the year for the USA and the UK.
In the recent years exponents $\delta<1$ are reported, indicating that large cities have smaller density than small ones.
}
\label{fig:delta}
\end{centering}
\end{figure}

While Eq.~(\ref{eq:easg}) describes the city emissions in terms of extensive quantities (city-wide totals), it has been argued that the intensive quantities (per capita rates) are a better representation \cite{ShaliziCR2011}.
Thus, next we discuss the dependence of \emph{per capita emissions} on the \emph{density of cities} \cite{UmSLJK2009}. 
Dividing by the population $s$, Eq.~(\ref{eq:easg}) leads to $\frac{e}{s}\sim \frac{s^\gamma}{s}$. 
Using the \emph{fundamental allometry} (population area allometry, see \cite{StewartJQ1947,BattyF2011,FluschnikKRZRKR2014} and references therein), 
\begin{equation}
s\sim a^\delta
\, ,
\label{eq:sadelta}
\end{equation}
where $a$ is the area of the city with population $s$, results in
$\frac{e}{s}\sim \frac{s^\gamma}{a^\delta}$.
Introducing the density $r=\frac{s}{a}$, we find the following proportionalities for the emissions per capita,
\begin{equation}
\frac{e}{s}\sim r^\delta s^{\gamma-\delta}
\end{equation}
or equivalently 
\begin{equation}
\frac{e}{s}\sim r^\gamma a^{\gamma-\delta}
\, .
\end{equation}
Thus, in general both quantities, density and population size, have an influence. 
Recovering Eq.~(\ref{eq:easg}), only for $\delta=\gamma$ and $\delta=1$ is there a pure dependency on the density and size, respectively.

In Fig.~\ref{fig:delta} we display $\delta$ exponents, the exponent of the relation between city area and population size, for the USA and UK as collected by \citeauthor{BattyF2011}~\citeyear{BattyF2011}~\cite{BattyF2011}.
As can be seen, the reported $\delta$-values decreases along time.
Qualitatively, this picture is congruent with Fig.~\ref{fig:slopes}, suggesting that as countries develop, large cities become less dense and more efficient in terms of emissions.
Assuming monotonous development of the GDP/cap along the years, the cross-sectional Fig.~\ref{fig:slopes} and the temporal Fig.~\ref{fig:delta} suggest similarity of the exponents, i.e.\ $\gamma\approx\delta$, emphasizing the importance of the density ($\frac{e}{s}\sim r^\gamma$). 
This means the higher CO$_2$ efficiency of large cities in developed countries could come along with lower densities of those cities.

It needs to be mentioned that scaling relations strongly depend on the definition of cities \cite{ArcauteHFYJB2014,LoufB2014EPB}.
Functional or morphological units can influence the resulting exponents, so we cannot exclude a bias if developing countries systematically use different definitions than the developed ones.
A similar argument could also apply for Fig.~\ref{fig:delta}. 
At least the most recent $\delta$ value for the USA corresponds to MSA, which are based on a functional definition -- previous measures are based on urban areas. 
Since functional definitions became more and more widely used after the 60s, this could also affect the trend observed for the UK.

\section{Country emissions}
\label{sec:countryemissions}

Relating emissions from cities to the emissions
from an entire country 
could help to understand the drivers of emissions.
Estimating the emissions of an entire country 
with the aforementioned relationship between city size and emissions, Eq.~(\ref{eq:easg}), requires taking into account 
how many cities exist for each size.
The size distribution of cities is described by the
Zipf-Auerbach law \cite{ZipfGK1949,AuerbachF1913,RybskiD2013}, which expresses that city sizes follow broad distributions, 
i.e.\ the probability density is proportional to $s^{-\zeta}$.
For cities it has been found that the Zipf-Auerbach exponent is $\zeta\simeq 2$, see e.g.\ \cite{RozenfeldRGM2011} and references therein.
In addition to the assumptions about the size distribution of cities 
we disregard emissions from non-urban regions.

Under these circumstances, 
the total urban emissions of a country are calculated by the sum of the number 
of cities of certain population size times the typical emissions of the corresponding city size.
We derive $E$, the expression for 
the urban emissions from the considered country (see Appendix~\ref{sec:app:deriv}), 
for $\zeta=2$
\begin{equation}
\label{eq:Esmax2limit}
E = d\,S\,f \hspace{1cm}{\rm with}\hspace{1cm}
f=
\left\{
  \begin{array}{lr}
    \frac{1}{\gamma -1} \frac{s_{\rm max}^{\gamma -1}-1}{\ln(s_{\rm max})} & 
      \mbox{if }\gamma\ne 1\\
    1 & \mbox{if }\gamma= 1 
  \end{array}
\right.
\, ,
\end{equation}
where $S$ is the total population, i.e.\ the sum over all city populations, and $d$ is the proportionality constant from Eq.~(\ref{eq:easg}).
Accordingly, for $\gamma\ne 1$ a factor $f$, depending on the population size of the largest city $s_{\rm max}$, comes into play capturing the deviations from the country population proportionality. 
\begin{itemize}
\item[$\gamma>1$:] The factor $f$ increases monotonously. 
Larger $s_{\rm max}$ imply relatively larger country emissions.
\item[$\gamma\simeq 1$:] $E$ is independent of $s_{\rm max}$ and 
it is proportional to the population of the country, 
i.e.\ $d=E/S$ are the per capita emissions. 
\item[$\gamma<1$:] The factor $f$ decreases monotonously. 
As the largest city population increases, the total emissions $E$ decrease relatively.
\end{itemize}
As indicated above, the exponent~$\gamma$ reflects the development of the country, and the special case $\gamma=1$ marks a transition made by economically emerging countries.
In any case, the proportionality constant $d$ and the population of the 
entire country $S$ remain.
The normalization done in Sec.~\ref{sec:cityemissions} levels out the allometric relations to $d\approx 1$.

However, so far we have not considered any relation between $S$ and $s_{\rm max}$. 
By doing so, the transition point, based on the country emissions can be further \emph{refined}. 
In Appendix~\ref{sec:largestcity} we depict a third scaling relation, namely between the total population of a country and the population of the largest city, i.e.\ $S=b\, s_{\rm max}^\tau$.
Eliminating $S$ in Eq.~(\ref{eq:Esmax2limit}), the transition is given by $\tau+\gamma=2$, i.e.\ at $\gamma\approx 0.8$ the change occurs from super-linear (for $\gamma>0.8$) to sub-linear (for $\gamma<0.8$) influence of the largest city (neglecting the case $\delta\ne 1$).

\section{Summary and discussion}
\label{sec:discussion}

In summary, we elaborate on the trinity of scaling relations of urban CO$_2$ emissions, consisting of 
(i) urban CO$_2$ emissions vs.\ city population size (exponent $\gamma$), 
(ii) city population vs.\ city area (fundamental allometry, exponent $\delta$), and 
(iii) population of largest city vs.\ population of country (largest city allometry, exponent $\tau$).
The analysis, based on our literature survey, indicates that $\gamma$ is non-universal and dependent on the GDP/cap of the corresponding country. 
A transition occurs at approximately $10,000$\,US\$, where the emissions exponent changes from $\alpha>1$ (for small GDP/cap) to $\alpha\le 1$ (for large GDP/cap). 
This implies that while in developing countries small cities are more CO$_2$ efficient, in developed countries this is the case for the large ones.
Thus, we cannot make any general statement regarding whether small or large cities perform favorably in relation to CO$_2$ efficiency.

In per capita terms, the GDP-dependence resembles an inverted U-shape, which is also known as Environmental Kuznets Curve, see \cite{KornhuberRCKSR2015} and reference therein.
It can be related to the interplay between sectoral composition and urbanization \cite{LutzSRKR2013}.
While developing countries mostly exhibit agrarian and increasingly industrial economies with small fraction of urban population, developed countries exhibit service and decreasing industrial sector economies with large fraction of urban population.

From simple considerations we obtain that, in general, the urban emissions depend on both population size and density. 
The latter is characterized by the fundamental allometry (ii).
Studying the corresponding exponents from \cite{BattyF2011} we find that $\delta$ seems to be time-dependent. 
Considering that the GDP/cap increases continuously over the years, we draw the analogy between the relations (i) and (ii) and hypothesize that gains of efficiency might be related to the density of the cities.
Unfortunately, from our literature survey only an insufficient number of density figures could be obtained.

Nevertheless, density scaling relations for urban CO$_2$ emissions are analyzed in \cite{Gudipudi2015}.
Regarding transport emissions, \citeauthor{NewmanK1989}~\citeyear{NewmanK1989}~\cite{NewmanK1989} report an exponential decrease of gasoline use per capita with increasing urban density. 
In Appendix~\ref{sec:NKFig1} we display the data of \cite{NewmanK1989} in double-logarithmic scale and fit a power-law leading to $\frac{e}{s}\sim r^{0.92\pm 0.07}$. 
This is contradicting our findings and the solution might lay in the superposition of size and density scaling, as derived in Sec.~\ref{sec:cityemissions}. 
Another reason could be a bias in the definition of the cities investigated.
Moreover, \citeauthor{LoufB2014SciRep}~\citeyear{LoufB2014SciRep}~\cite{LoufB2014SciRep} derive that the emissions per capita are not a simple function of density, but rather of surface area.

With the aim of characterizing country-wide urban emissions, we theoretically integrate the city emissions (assuming Zipf-Auerbach law with $\zeta\simeq 2$) and find that for the general case $\gamma\ne 1$ the population size of the largest city in a country plays an important role. 
Depending on the value of $\gamma$ it leads to relatively increased ($\gamma>1$) or reduced ($\gamma<1$) country-wide urban emissions.
Interestingly, according to the scaling relation (iii) with $\tau\approx 1.2$ the above mentioned transition point can be readjusted to $\gamma\approx 0.8$ when considering the country-wide emissions as a function of the size of the largest city are considered. 
To our knowledge the scaling relation (iii) is previously unknown.

Coming back to the initially posed question about the sustainability of cities, we first need to confine the term "sustainability" to the climate change context and CO$_2$ emissions. 
From our analysis we can only give a differentiated answer, namely that urbanization drives climate change in developing countries and mitigates climate change in developed ones. 
However, our findings suggest that density-scaling might as well play a role, and we therefore call for a coherent analytical and empirical analysis that takes into account both the population size \emph{and} the population density.

\appendix

\begin{widetext}

\section{Derivation of Eq.~(\ref{eq:Esmax2limit})}
\label{sec:app:deriv}

According to the Zipf-Auerbach law the probability density of city sizes is 
proportional to $s^{-\zeta}$ for $1\le s\le s_{\rm max}$,
where $\zeta$ is the Zipf-Auerbach exponent and $s_{\rm max}$ is the population of the largest city.

For the calculation of the total emissions of a country we need the 
\emph{absolute distribution function}
\begin{equation}
p(s)=c \, s^{-\zeta} \qquad (1\le s\le s_{\rm max})
\, ,
\label{eq:app:totalzipf}
\end{equation}
where $p(s)$ is the absolute number of cities of size $s$. 
We can calculate the constant $c$ with the following identity
\begin{equation}
S = \int_1^{s_{\rm max}} p(s)\,s\,{\rm d}s = 
\int_1^{s_{\rm max}} c\,s^{-\zeta+1}\,{\rm d}s
\, ,
\end{equation}
where $S$ is the population of the entire country, 
leading to
\begin{equation}
c = \frac{S}{\int_1^{s_{\rm max}} s^{-\zeta+1}\,{\rm d}s}
\label{eq:app:tildec}
\, .
\end{equation}

The total emissions of a country are given by the integral over the product of 
the absolute distribution of city sizes, Eq.~(\ref{eq:app:totalzipf}), 
and the emissions of the considered city sizes, Eq.~(\ref{eq:easg}),
\begin{equation}
E=
\int_1^{s_{\rm max}} p(s)\,e(s){\rm d}s
=
\int_1^{s_{\rm max}} c\,s^{-\zeta}\,d\,s^\gamma{\rm d}s
\, ,
\end{equation}
where $E$ are the total emissions of a country.

Using Eq.~(\ref{eq:app:tildec}) and taking constants outside
the integral gives
\begin{equation}
E \enspace = \enspace 
\frac{d\,S}{\int_1^{s_{\rm max}} s^{-\zeta+1}\,{\rm d}s}
\int_1^{s_{\rm max}} s^{\gamma-\zeta}{\rm d}s
\, .
\end{equation}
Assuming $\zeta\simeq 2$ we obtain
\begin{equation}
E \enspace = \enspace 
\frac{d\,S}{\int_1^{s_{\rm max}} s^{-1}\,{\rm d}s}
\int_1^{s_{\rm max}} s^{\gamma-2}{\rm d}s
\enspace = \enspace 
\frac{d\,S}{\ln(s_{\rm max})}
\int_1^{s_{\rm max}} s^{\gamma-2}{\rm d}s
\, .
\end{equation}
In order to integrate, one needs to distinguish two cases.

\begin{figure*}
\begin{centering}
\includegraphics[width=0.75\textwidth]{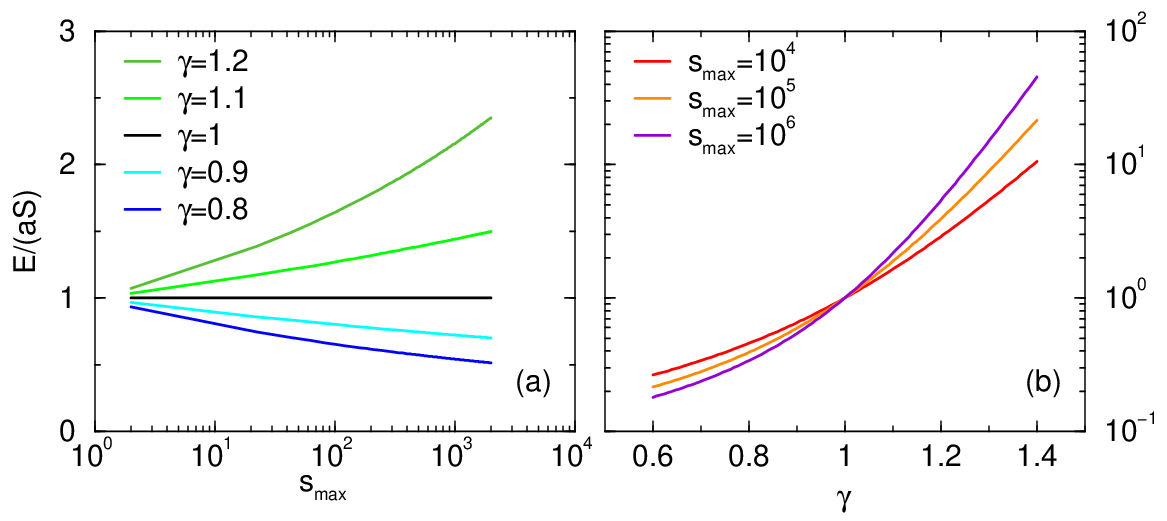}
\caption{
Illustration of theoretical country-wide urban emissions.
We plot the rescaled emissions $E/(dS)$ (a) as a function of $s_{\rm max}$ and (b) as a function of $\gamma$.
The curves are given by Eq.~(\ref{eq:Esmax2limit}) with $\zeta=2$.
}
\label{fig:illu}
\end{centering}
\end{figure*}

\begin{enumerate}
\item 
$\gamma=1$ leads to
\begin{equation}
E = \frac{d\,S}{\ln(s_{\rm max})}\ln(s_{\rm max}) 
=d\,S
\, .
\label{eq:gz1}
\end{equation}
This means the country emissions are determined by the constants $d$ and
$S$, as expected, since for $\gamma=1$ the constant $d$ corresponds to the per capita emissions.
\item 
$\gamma\ne 1$ leads to
\begin{equation}
E = 
\frac{d\,S}{\ln(s_{\rm max})}
\frac{s_{\rm max}^{\gamma-1}-1}{\gamma-1}
\, .
\label{eq:gnez1}
\end{equation}
Again, $E$ is determined by $d$ and $S$, but in this case 
there is a power-law term and a logarithmic term of $s_{\rm max}$ 
as well as a term depending on $\gamma$. 
\end{enumerate}
The cases $\gamma>1$, $\gamma=1$, and $\gamma<1$ are illustrated in Fig.~\ref{fig:illu}, where $E/(dS)$ is plotted as a function of $s_{\rm max}$ and as a function of $\gamma$.

\end{widetext}

\section{Correlations between the population of a country and 
the population of its largest city}
\label{sec:largestcity}

\begin{figure*}
\begin{centering}
\includegraphics[width=0.75\textwidth]{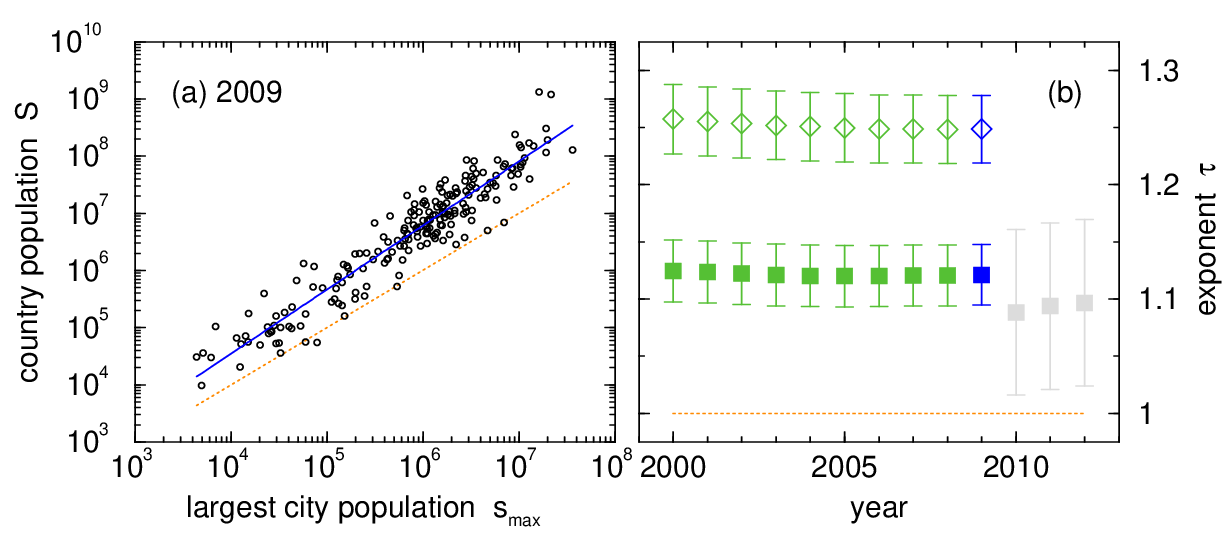}
\caption{
Correlations between the population of a country and 
the population of its largest city.
(a) Country population $S$ versus largest city population $s_{\rm max}$ 
for 205 countries in the year 2009. 
Each circle represents one country, the solid blue line is a regression
according to Eq.~(\ref{eq:smaxs}), and the dotted orange line is 
a guide for the eye with slope $1$. 
(b) The exponent $\tau$ in Eq.~(\ref{eq:smaxs}) was obtained from 
regressions as in panel (a) for various years (full squares). 
Open diamonds represent the $\tau$ values obtained from fitting 
the inverse function of Eq.~(\ref{eq:smaxs}) \cite{FluschnikKRZRKR2014}. 
The error-bars indicate the standard errors from fitting.
The colors distinguish the number of available countries: 
green, 204; blue, 205; and grey approximately 120 (mostly large countries). 
The dotted orange line indicates $\tau=1$. 
The exponent $\tau$ is clearly above $1$, 
i.e.\ $\tau\approx 1.2$, and minimally decreases over time, 
whereas the trend is smaller than the statistical error.
Data: \cite{Worldbank2013a,Worldbank2013b}.
}
\label{fig:smaxs}
\end{centering}
\end{figure*}

In Fig.~\ref{fig:smaxs} we find power-law correlations 
between the country populations, $S$, and the corresponding largest city 
populations, $s_{\rm max}$, according to
\begin{equation}
\label{eq:smaxs}
S=b\, s_{\rm max}^\tau
\, ,
\end{equation}
with $\tau\approx 1.2$ and $b\simeq 1$. 
A similar relation has been studied between urban population of the country and population of metropolis \cite{PumainME1997} -- the reported exponent $1/0.81\approx 1.23$ is close to our value.

This relation can be used to eliminate the country population, $S$, 
in Eq.~(\ref{eq:Esmax2limit}) and the transition between super- and sub-linear scaling with $s_{\rm max}$ is then given by $\tau+\gamma=2$, i.e. at $\gamma\approx 0.8$ a change occurs from super-linear (for $\gamma>0.8$) to sub-linear (for $\gamma<0.8$) influence of the largest city (neglecting the case $\delta\ne 1$).

\section{Emissions versus density curve by Newman and Kenworthy}
\label{sec:NKFig1}

\begin{figure*}
\begin{centering}
\includegraphics[width=\columnwidth]{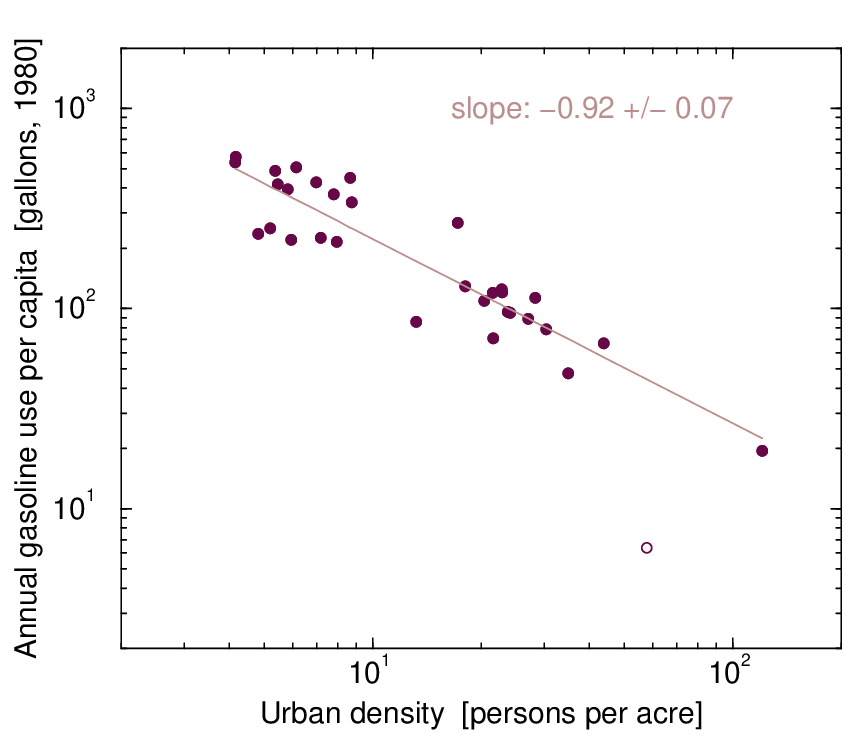}
\caption{
Gasoline use per capita versus population density (1980) as extracted from \cite{NewmanK1989}. 
The power-law fit -- disregarding Moscow (open circle) -- leads to the slope $0.92\pm 0.07$.
In contrast to Fig.~1 in \cite{NewmanK1989}, here the data is plotted in double-logarithmic scale. 
}
\label{fig:NKFig1}
\end{centering}
\end{figure*}

In Fig.~\ref{fig:NKFig1} the values of annual gasoline use per capita as a function of urban density as extracted from Fig.~1 in \cite{NewmanK1989} are plotted in double-logarithmic scale.
While the authors report an exponential relationship, we fit a power-law and obtain a slope $0.92\pm 0.07$ (disregarding the point for Moscow). 
This result might be superposed by a size-effect as discussed in Sec.~\ref{sec:cityemissions}.

\begin{acknowledgments}
We thank A.~Nockert and S.~Huber for help with the data as well as 
N.~Schwarz, X.~Gabaix, R.~Gudipudi, R.~Louf and several anonymous reviewers for useful comments.
The research leading to these results has received funding from the European
Community's Seventh Framework Programme under Grant Agreement No.~308497
(Project RAMSES).
We appreciate financial support by the Federal Ministry for Education and
Research of Germany who provided support under the rooftop of the \emph{Potsdam
Research Cluster for Georisk Analysis, Environmental Change and Sustainability}
(PROGRESS) Initiative (grant number \#03IS2191B).
\end{acknowledgments}

\begin{widetext}



%

\end{widetext}

\end{document}